\documentclass[12pt,preprint]{aastex}

\shorttitle{SiO masers in the GC}
 \shortauthors{Li et al.}

\begin{document}

\title{ATCA observations of SiO masers in the Galactic center }

\author{Juan Li\altaffilmark{1,2,3}, Tao An\altaffilmark{1}, Zhi-Qiang Shen\altaffilmark{1,2}, Atsushi Miyazaki\altaffilmark{4}}

\altaffiltext{1}{Shanghai Astronomical Observatory, Chinese Academy
of Sciences, Shanghai 200030, China; lijuan@shao.ac.cn;
antao@shao.ac.cn, zshen@shao.ac.cn} \altaffiltext{2}{Joint Institute
for Galaxy and Cosmology (JOINGC) of SHAO and USTC, Shanghai 200030,
China;}
 \altaffiltext{3}{Department of Astronomy, Nanjing University, Nanjing 210093, China;}
\altaffiltext{4}{Mizusawa VLBI Observatory, National Astronomical Observatory of Japan, Mizusawa, Oshu, Iwate 023-0861, Japan; amiya@miz.nao.ac.jp}

\begin{abstract}

We present the Australia Telescope Compact Array (ATCA) observations
of the SiO  masers in the Galactic center in transitions of v=1,
J=2--1 at 86~GHz and v=1, J=1--0 at 43~GHz. Two 86-GHz SiO masers
were detected within the central parsec, and they are associated
with IRS 10EE and IRS 15NE, respectively. We detected eighteen
43-GHz SiO masers within a projected separation of $\lesssim$2 pc
from Sagittarius A* (Sgr A*), among which seven masers are newly
discovered from our observations. This raises the total number of
43-GHz SiO masers within the central 4 parsecs of the GC region to
22. Simultaneous observations at 86 and 43~GHz showed that the
intensity of 43-GHz SiO maser is $\sim$3 times higher than that of
86-GHz maser in IRS 10EE (an OH/IR star), while the integrated flux
of the SiO maser emission at 43 GHz is comparable with that at
86~GHz in IRS~15NE (an ordinary Mira variable). These results are
consistent with previous observations of massive late-type stars in
the Galaxy in which the 86-GHz SiO maser is in general weaker than
the 43-GHz SiO maser in OH/IR stars, while the two transitions are
comparably strong in Mira stars.

\end{abstract}


\section{Introduction}
\qquad

SiO masers located in the extended circumstellar envelopes of
late-type giant stars are believed to be reliable tracers of the
Galactic dynamics since they can be treated as point-like objects
and are not subject to non-gravitational forces such as magnetic
fields or stellar wind collisions. A number of SiO maser surveys
toward the Galactic center (GC) region have been carried out at 43
GHz in order to study the structure and dynamics of the stellar disk
in the Inner Galaxy (Izumiura et al. 1998; Miyazaki et al. 2001;
Deguchi et al. 2000, 2002; Sjouwerman et al. 2002, 2004; Imai et al.
2002). So far the most sensitive search for 43-GHz SiO masers in the
central few parsec region of the GC was made with the VLA and VLBA,
and the accurate position determination of the SiO masers has been
used to calibrate the astrometry in the infrared images of the same
region with an accuracy of $\sim$1 mas (Menten et al. 1997; Reid et
al. 2003, 2007; Oyama et al. 2008). Moreover fifteen SiO maser
sources have been observed with a projected separation
$<50^{\prime\prime}$ from Sgr A* at 43 GHz; lower limit of the
dynamic mass enclosed within a given radius was derived from the
3-dimensional velocity of the maser star. Whereas only in IRS~9, the
derived enclosed mass limit is $\sim$20 percent higher than the mass
of known objects in the GC, i.e., the supermassive black hole (SMBH)
and the luminous stars (Reid et al. 2007). In fact previous VLA
observations of SiO masers were constrained by the limited velocity
coverage of the receiver band. Searching for high-velocity SiO maser
stars is critical for exploring the mass distribution in the GC.

At 86 GHz, Lindqvist et al. (1991) searched for SiO maser line
emission in a sample of 31 OH/IR stars close to the GC using the
Nobeyama 45m telescope, but they failed to obtain distinctive
detection of any 86 GHz masers. Messineo et al. (2002) detected 271
SiO masers in a sample of 441 late-type stars in the Inner Galaxy
from their high-sensitivity IRAM-30m observations. Limited by the
low angular resolution and sensitivity of single-dish observations,
the number and spatial distribution of SiO maser sources within a
few parsecs of the GC remain uncertain. Until recently,
high-angular-resolution observations of 86-GHz SiO maser line
emission from the GC have been made possible using the Australia
Telescope Compact Array (ATCA) of the Australia Telescope National
Facility (ATNF) (this paper and L.O. Sjouwerman et al., private
communication).

In addition to the important role on accurate astrometry, SiO maser
is also an excellent probe of the physical condition in the
circumstellar envelope. VLBI observations of SiO masers in massive
late-type stars distributed in the Galactic disk showed that these
maser spots are located in the innermost ($\sim$4--8 AU) regions of
the circumstellar envelopes of the stars (Diamond \& Kemball 2003),
whereas the pumping mechanism of the SiO maser line is still an open
question (Doel et al. 1995; Desmurs et al. 2000; Humphreys et al.
2002). The SiO masers are strongly variable over timescales of
$\sim$1 yr, therefore simultaneous observations of the GC SiO masers
at multiple transitions would provide helpful information for
constraining the excitation conditions of the SiO maser lines under
extreme astrophysical environment.

In this Letter, we present the results from the ATCA observations of
SiO maser lines at two transitions (v=1, J=2--1 (86 GHz) and v=1,
J=1--0 (43 GHz)) within a few parsecs from Sgr~A*. The ATCA
observations and data reduction are described in Section
\ref{observation}, and the results and analysis are presented in
Section \ref{result}. In Section \ref{summary} we summarize the
observational results. Throughout this paper we adopt a GC distance
of 8 kpc (Reid 1993) and a SMBH mass of $4\times 10^{6}M_{\odot}$ at
the GC (Ghez et al. 2008; Gillessen et al. 2009).

\section{OBSERVATIONS AND DATA REDUCTION}
\label{observation}

We carried out ATCA observations of the Galactic center at 86 and 43
GHz in five sessions (2006 June 9, 2006 August 13, 2007 October 21,
2008 June 3-4 and 2008 October 3). These observations were initially
made with the purpose of investigating the variability of Sgr A* at
millimeter wavelengths (Li et al. 2009). The observations on 2008
October 3 were made quasi simultaneously at 43 and 86 GHz by
interchanging the receivers at the two frequencies with a duty cycle
of about 20 minutes. The observations on other dates were performed
solely at a single frequency. Except for the observation on 2008
June 3-4, other observations were made under good weather condition.
Table 1 summarizes the observational parameters.

Quasar 3C 279 was observed at the beginning of each observing run to
calibrate the bandpass. The observations were pointed at Sgr A*,
allowing for detection of SiO masers located in the primary beam of
the ATCA antenna, i.e., $\theta_{FWHP}$ of $\sim 36^{\prime\prime}$
and $\sim 72^{\prime\prime}$ at 86 and 43 GHz, respectively. The
pointing were checked and corrected every half an hour by observing
either a nearby quasar PKS~1730-130 (NRAO~530) or a strong SiO maser
VX~Sgr. Quasar PKS~1741-312 and SiO maser OH 2.6--0.4 were
interlaced with Sgr~A* to track the elevation-dependent gains.

The data were reduced using the {\it MIRIAD}
package\footnote{http://www.atnf.csiro.au/computing/software/miriad/}.
We first calibrated the bandpass and elevation-dependent complex
gains following the standard procedure described in the {\it MIRIAD}
Cookbook. Next we derived the flux density scale for the visibility
data observed in 2007 and 2008 using the planet Uranus. The
resulting uncertainty of the absolute flux density was
$\lesssim20\%$. For the observations in 2006, the Uranus was
resolved, therefore the flux density scale was determined from the
quasar NRAO~530 on 2006 June 9 (2.27 Jy from the ATCA calibrator
monitoring program) and from PKS 1921-293 (8.44 Jy) on 2006 August
13, respectively. Then we made a continuum map of the Galactic
center from the visibilities on line-free channels and further
carried out a few iterations of phase-only self-calibration to
correct residual phase errors. The gain solutions derived from the
continuum data were transferred and applied to the line data. Using
{\it MIRIAD} task {\rm UVLIN}, the continuum level was determined by
a linear fitting to the line-free channels and was subtracted from
the calibrated line data in the {\it uv} plane. A spectral-line cube
was constructed from mapping the continuum-subtracted data. In order
to increase the sensitivity in the channel maps, we have smoothed
the line data to a velocity resolution of 1 km s$^{-1}$ at 86 GHz
and 2 km s$^{-1}$ at 43 GHz. The typical {\it rms} noise in
individual channels is 5 mJy at 43 GHz, and 13--55 mJy at 86~GHz,
depending on the weather condition and effective integration time
(see Table 1).

We searched for SiO maser sources within a radius of
$25^{\prime\prime}$ and $50^{\prime\prime}$ centering at Sgr A* at
86 and 43 GHz, respectively. For those SiO maser sources detected
above 5$\sigma$, we derived their positions in two steps: first, we
fitted their brightness distribution with a two-dimensional Gaussian
using the task {\rm IMFIT} in each channel; then, we made
variance-weighted averaging to obtain its mean position and the
uncertainty.

\section{RESULTS AND ANALYSIS}
\label{result}

\subsection{86 GHz masers}

Two SiO masers were detected in the 86 GHz images with peak flux
densities higher than 5$\sigma$. These two sources were identified
to be associated with IRS~10EE and IRS~15NE, based on the consistent
positions and radial velocities determined from previous
observations (Reid et al. 2003, 2007). IRS~10EE appeared at four
epochs ($\sim4\sigma$ on 2008 October 3) and displayed strong
variability on timescale of months. IRS~15NE was only distinctively
detected on 2007 October 21. It was marginally detected on 2008 June
4 ($\sim3\sigma$) as a result of high \emph{rms} noise level. The
late-type giants and supergiants in the GC are likely detectable
over a wide radial velocity range of $\pm$350 km s$^{-1}$ (e.g.,
Figer et al. 2003). However, current 86-GHz observations were
constrained by the limited velocity coverage and relatively low
sensitivity. The positions, peak flux densities and
velocity-integrated flux densities of IRS~10EE and IRS~15NE are
listed in Table 2. The {\it rms} noise of the integrated flux
density is estimated as $\triangle I = \sigma_{rms}\triangle v
(N_{chan})^{1/2}$, where $\sigma_{rms}$ is the {\it rms} noise level
in individual channel maps, $\triangle v$ is the velocity
resolution, and $N_{chan}$ is the number of channels over which the
$\Delta I$ is calculated (Klaassen \& Wilson 2007).

We will comment on these two 86-GHz maser sources as follows.

(1) {\it IRS 10EE} was identified as an OH/IR star and has been
observed with SiO, OH and H$_2$O maser line emission (Lindqvist et
al. 1990, 1992; Menten et al. 1997;  Sjouwerman et al. 2002; Reid et
al. 2003, 2007; Oyama et al. 2008; Peeples et al. 2007). In infrared
(IR) bands it was classified as a long-period variable, and
exhibited variability on various timescales ranging from several
days to a few months (Tamura et al. 1996; Wood et al. 1998; Ott et
al. 1999; Peeples et al. 2007). X-ray emission was detected within
0.$^{\prime\prime}$8 of IRS 10EE (Muno et al. 2004) and a binary
system was proposed to explain its relatively high X-ray luminosity
as an AGB star (Peeples et al. 2007). The binary scenario of
IRS~10EE was further supported by the VLBA monitoring of its proper
motion (Oyama et al. 2008). Spectra of 86- and 43-GHz SiO maser
lines of IRS 10EE from our observations are shown in the upper panel
of Figure 1. They indicated a flux density variation on timescale of
months, in agreement with previous radio and IR observations.
Long-term monitoring of SiO maser emission in IRS~10EE and
comparison with the optical phase change would provide an important
tool for understanding the pumping mechanisms in this source (Pardo
et al. 2004).

(2) {\it IRS 15NE} showed not only strong characteristic cool star
features, but broad He I and Br$\gamma$ emission in the IR spectrum,
suggesting that it is a close blend (Genzel et al. 1996, 2000; Blum
et al. 2003, Oyama et al. 2008). VLBA observations of IRS~15NE at 43
GHz showed that the maser source was only slightly resolved and
extended with a diameter of about 2 mas (Oyama et al. 2008). This
SiO maser source is relatively weak compared with IRS~10EE. It is
variable at both 43 and 86 GHz (lower panel in Figure 1). The 43 GHz
peak flux density on 2008 June 3 was only a seventh of that on 1998
observing epoch of Reid et al. (2003). The velocity of the peak
86-GHz maser feature changed from about $-$11.7 km s$^{-1}$ on 2007
October 21 to about $-$13.0 km s$^{-1}$ on 2008 June 4. For a star
at a projected separation $R$ from the GC, the radial-direction
component of the acceleration, $a_r$, due to the gravitation of the
enclosed mass $M(r)$ can be calculated as:
\begin{equation}
 \frac{a_r}{[km \,s^{-1}\, yr^{-1}]} = \frac{GM(r)\sin\alpha}{(\frac{R}{{\rm cos}\alpha})^2}  \leq 0.05
(\frac{\theta}{10''})^{-2}(\frac{M(r)}{4 \times 10^6 M_{\odot}})
\end{equation}
where $G$ is the gravitational constant,
$\alpha$ is the angle between the radius vector to the star and the
plane of the sky containing the central mass (0$^\circ$$\leq \alpha
\leq$90$^\circ$), and $\theta$ is the observed angular separation.
The $a_r$ gets its maximum value in the case of $\alpha=35\degr$.
Calculations based on the observed parameters of IRS~15NE suggest
that the acceleration caused by the enclosed mass is not significant
and far lower than the rate of the peak velocity variation. On the
other hand, IRS~15NE seems to be a special case showing large
variation of the peak velocity at the 43 GHz (Table 2). Therefore,
the observed variation of the peak velocity of IRS 15NE seems
unlikely to be caused by the gravitation acceleration by the dynamic
center of the GC. The resolution of current observations is not high
enough to explore whether the changes of the peak velocity are
related to the variation in the spatial distribution of the masering
regions.

\subsection{43 GHz masers}

Seventeen 43-GHz SiO masers were identified on 2008 June 3 based on
both radial velocity and position information. Among them eleven
sources have been observed previously (Menten et al. 1997; Reid et
al. 2003, 2007) and six are newly detected. One of the six new SiO
masers,  offset (+0.$^{\prime\prime}$9, $-8.^{\prime\prime}$0) from
Sgr~A*, was associated with IRS 14NE, a known AGB star (Blum et al.
1996; Ott et al. 1999; Peeples et al. 2007). Following the
nomenclature of the GC SiO masers adopted by Reid et al. (2007), the
other five new SiO masers were named as SiO-18 ($-77.6$ km
s$^{-1}$), SiO-19 ($-29.4$ km s$^{-1}$), SiO-20 ($-20.3$ km
s$^{-1}$), SiO-21 ($+13.8$ km s$^{-1}$) and SiO-22 ($+33.6$ km
s$^{-1}$) in a sequence of ascending velocity.

We detected fourteen 43-GHz SiO masers on 2008 October 3, among
which eleven masers were in agreement with previous observations
(Menten et al. 1997; Reid et al. 2003, 2007) and two masers (SiO-18
and SiO-22) were also found on 2008 June 3. The remaining one maser
($V_{LSR}=+80$ km s$^{-1}$) was found at ($+24.\arcsec0$,
$+20.\arcsec3$) northeast of Sgr~A* and represents a new detection.
We named it as SiO-23.

These detections at both 86 and 43 GHz are tabulated in Table 2.
Together with previous observations, our new detections of SiO
masers in the GC raise the total number of the 43-GHz SiO masers in
the central 100\arcsec{} (or 4 pc) region to 22, increasing by
$\sim$50 percent. Figure 2 shows composite spectra of the 43-GHz SiO
masers derived from two observations in 2008. Figure 3 displays a
sketch map of the 2-dimensional distribution of 43-GHz SiO masers in
the inner 100\arcsec{} region of the GC. The contours in Figure 3
represent the continuum emission at 43 GHz. It is interesting to
note that almost all sources with negative radial velocities have
projected separations away from Sgr~A* less than 25$^{\prime\prime}$
($\sim$1 pc); while most sources with positive radial velocities are
located at larger projected distances $>25\arcsec$. The origin for
such a velocity distribution of the GC SiO maser stars is not clear.
The IR observations of the massive stars in the GC suggest that the
red giants and supergiants are belonging to a relax system and show
isotropic 3-dimensional velocity distribution. Was the systematic
change of the radial velocities of the SiO maser stars related to an
intrinsically extraordinary property of the SiO maser stars
(different from those non-maser red giants) or just a selection
effect? More observations using the ATCA and EVLA with receiver
bandwidth that covers the full velocity range of the GC SiO masers
are necessary to obtain a complete view of the radial velocity and
spatial distribution of SiO masers in the GC region and thus to
solve this velocity distribution puzzle.

\subsection{Relative Strength of SiO Maser transitions}

Figure 1 compares the SiO maser line spectra of IRS~10EE from the
quasi-simultaneous observations at 86 and 43 GHz on 2008 June 4 and
2008 October 3, and the quasi-simultaneous spectra of IRS~15NE at 86
and 43 GHz on 2008 June 4.  The 43-GHz integrated flux of IRS 10EE
is about 3-5 times higher than that of 86 GHz in the two
observations. Different from IRS 10EE, the 43 GHz integrated flux of
IRS 15NE is comparable with that of 86 GHz. The different behaviors
of the SiO maser line emission in IRS~10EE (an OH/IR star) and
IRS~15NE (an ordinary Mira) seem to be consistent with previous
observations of massive late-type stars in the Galaxy (Nyman et al.
1986, 1993; Lindqvist et al. 1992) which showed that the 86-GHz SiO
maser emission was weak in OH/IR stars compared with ordinary Mira
variables. The high mass loss rate in OH/IR stars results in much
turbulence to quench the 86-GHz maser excitation, as provides a
possible explanation for relatively weak 86-GHz SiO maser emission
in OH/IR stars. The radiative pumping should yield similar 43- and
86-GHz SiO intensities over broad ranges of SiO densities and
abundances, while the collisional pumping is sensitive to the SiO
column density and may give rise to different spatial distribution
and different intensities at 86 and 43 GHz (Phillips et al. 2003 and
references therein). Although the low-resolution observations in the
present paper are not allowed for inspecting the spatial
correspondence of 43- and 86-GHz maser spots, the large discrepancy
between the integrated flux of IRS 10EE at the two frequencies would
suggest that the radiative pumping might not be the dominant
mechanism in this source. On the other hand, the turbulent outflows
in IRS~10EE make it difficult for the circumstellar envelope to
sustain at a density of $(4-6)\times10^9$ cm$^{-3}$, which is a
scope for the 86-GHz maser emission to be effectively amplified
(Doel et al. 1995). For IRS 15NE, either radiative or collisional
pumping may be responsible for the observed 43- and 86-GHz SiO maser
emission. In order for the collisional mechanism to be at work for
IRS~15NE, the H$_2$ gas density in most regions of the envelope is
required to be close to a critical density $\sim5\times10^9$
cm$^{-3}$ (Doel et al. 1995). Long-term monitoring of the maser line
and infrared emissions from the GC SiO maser stars might be helpful
for understanding the connection between the SiO maser variability
and the IR stellar pulsation cycle.

\section{Conclusion}
\label{summary}

A well-distributed sample of SiO maser stars in both spatial and
velocity spaces is crucial for revealing the enclosed mass as a
function of the projected radius in the GC. We have made sensitive
search for SiO masers in the central 4-pc region of the GC at 43 and
86 GHz using the ATCA. The observations resulted in a detection of
two 86-GHz SiO masers and eighteen 43-GHz masers, of which seven are
discovered for the first time. Combined with previous observations
(Reid et al. 2007), the total number of the 43-GHz SiO masers in GC
region has been raised to 22. The current observations of the GC SiO
masers mainly made with the ATCA and VLA are concentrated in the
radial velocity range of $\pm$100 km s$^{-1}$. The updated
broad-band observing facilities such as the ATCA (with the CABB) and
the EVLA are excellent tools to gain a complete view of the
distribution of GC SiO masers. In particular, monitoring the
high-radial-velocity maser stars using the VLBI is especially
important for constraining the mass distribution in the GC.

In addition, the relative strength of SiO masers between 43 and 86
GHz from the simultaneous observations of IRS 10EE tends to favor a
collisional pumping mechanism in IRS 10EE. For IRS 15NE, either
radiative or collisional pumping mechanism, or both, might be at
work.

\acknowledgments

The Australia Telescope Compact Array is part of the Australia
Telescope which is founded by the Commonwealth of Australia for
operation as a National Facility managed by the CSIRO.

This work was supported in part by the National Natural Science
Foundation of China (grants 10625314, 10633010 and 10821302), the
CAS/SAFEA International Partnership Program for Creative Research
Teams and the Knowledge Innovation Program of the Chinese Academy of
Sciences (Grant No. KJCX2-YW-T03), and sponsored by the Program of
Shanghai Subject Chief Scientist (06XD14024) and the National Key
Basic Research Development Program of China (No. 2007CB815405,
2009CB24903). We thank Willem Baan for proof-reading and
constructive comments on the manuscript. JL would like to thank
helpful discussions with Lorant Sjouwerman on the ATCA observations,
and Xi Chen, Yongjun Chen, Bing Jiang and Zhiyu Zhang for comments
and suggestions on this work.


\clearpage

\begin{deluxetable}{c|c|c|c|c|c|c|c|l}
\centering
  \tablewidth{0pt}
  \tabletypesize{\scriptsize}
  \tablecolumns{9}
  \tablecaption{Summary of ATCA Observations }
 \tablehead{  Date      &    BW  &  $\triangle \nu^{a}$  & $\triangle v^{b}$ & $\triangle V^{c}$  & Beam   & $\triangle \tau$   & $\sigma$$^{d}$    & Phase Center      \\
           &     (MHz)  &     (kHz)  &  (km s$^{-1}$)          &  (km s$^{-1}$)     &   ($^{\prime\prime} \times ^{\prime\prime}$)   & (min)   & (mJy)      &      \\
}
\startdata
\multicolumn{9}{c}{86~GHz}  \\
            \hline
2006 June 9 &  16     &  62.5    & 0.22            &   $-$31 $\sim$ $+$22  &  $2.2 \times 1.6$  & 78  & 47      & RA: 17:45:40.045     \\
           &        &       &                 &                  &                  &    &        & DEC: $-$29:00:27.90     \\
2006 August 13 &  16     &  62.5    & 0.22            &   $-$31 $\sim$ $+$22  &  $2.3 \times 1.6$  &135  & 26      & as above  \\
2007 October 21  &  32    &   250    & 0.87             & $-$60 $\sim$ $+52$    &   $2.2 \times 1.6$  & 182  & 13      &  as above   \\
2008 June 4  &  32     &  250    & 0.87            & $-$60 $\sim$ $+$52     &   $5.9 \times 1.2$  & 58  & 48    &  RA: 17:45:40.038   \\
           &        &       &                 &                  &                   &   &       & DEC: $-$29:00:28.07  \\
2008 October 3  &  32    &   250    & 0.87             & $-$60 $\sim$ $+$52     &   $3.1 \times 1.9$  &  33   & 55      &   as above  \\
 \hline
\multicolumn{9}{c}{43~GHz}    \\
            \hline
2008 June 3   &  32    &   250    & 1.74             & $-$116 $\sim$ $+$110     &   $8.4 \times 2.3$  & 250  & 4    &  RA: 17:45:40.038   \\
           &        &       &                 &                  &                      &  &     & DEC: $-$29:00:28.07  \\
2008 October 3  &  32    &   250    & 1.74             & $-$116 $\sim$ $+$110     &   $6.3 \times 3.6$ & 29   & 5      &   as above  \\
\hline
 \enddata
  \tablenotetext{a}{: channel width in unit of kHz; $^b$: velocity resolution
corresponding to the channel width $\Delta \nu$; $^c$: velocity
coverage; $^d$: channels were binned with a width of 1 km s$^{-1}$
at 86 GHz and 2 km s$^{-1}$ at 43 GHz while mapping the SiO maser
line emission, then the {\it rms} noise in individual channels was
estimated.}
 \end{deluxetable}

\begin{deluxetable}{c|c|c|c|c|c|c|l}
\centering
  \tablewidth{0pt}
  \tabletypesize{\scriptsize}
  \tablecolumns{8}
  \tablecaption{86 and 43 GHz SiO maser sources detected in the central 4 parsecs of the GC}

 \tablehead{ SN & Name & $v_{LSR}$    & $\bigtriangleup \Theta_x$  & $\bigtriangleup \Theta_y$ & Peak   & Intensity  & Date \\
   &   &  (km s$^{-1}$)             & ($^{\prime\prime}$)                  &  ($^{\prime\prime}$)         &  (mJy) & (mJy km s$^{-1}$)  &
 }
 \startdata
 \multicolumn{8}{c}{86~GHz}    \\
 \hline
1 & IRS 10EE &  $-$27.7     & 7.74$\pm$0.10   &  4.23$\pm$0.40    &  290$\pm$20  &  470$\pm$80   & 2006 June 9 \\
  &          &  $-$27.4     & 7.70$\pm$0.01   &  4.22$\pm$0.04    & 380$\pm$20   & 690$\pm$40    & 2006 August 13 \\
  &          &  $-$27.5     & 7.68$\pm$0.08   &  4.33$\pm$0.06    &  340$\pm$10  &   700$\pm$30  & 2007 October 21 \\
  &          &  $-$27.6     & 7.62$\pm$0.12   &  4.44$\pm$0.49    & 450$\pm$20   &  610$\pm$100  & 2008 June 4 \\
  &          &  $-$27.5     & 7.67$\pm$0.71   &  4.26$\pm$0.50    & 240$\pm$4    &  350$\pm$110  & 2008 October 3 \\

2 &  IRS 15NE&  $-$11.7     &  0.99$\pm$0.18  &  11.09$\pm$0.22    & 139$\pm$3    &  392$\pm$26    & 2007 October 21 \\
  &          &  $-$13.0     &  1.13$\pm$0.08  &   9.60$\pm$0.49    & 182$\pm$11   &   297$\pm$68  & 2008 June 4 \\
         \hline
\multicolumn{8}{c}{43~GHz}  \\
 \hline
1 & SiO-18$^{*}$     &  $-$77.6     &  $-$18.88$\pm$0.03  & $-$25.80$\pm$0.14     & 39.6$\pm$0.8   & 72.5$\pm$8.0  & 2008 June 3  \\
  &                 &  $-$78.0     &  $-$18.70$\pm$0.09  & $-$25.16$\pm$0.07     & 22.9$\pm$0.3   & 44.7$\pm$10.0  & 2008 October 3  \\
2 & IRS 12N         &  $-$63.2     &  $-$3.41$\pm$0.04 & $-$6.28$\pm$0.19    & 283.7$\pm$3.0 &  5126.0$\pm$17.9 & 2008 June 3 \\
  &                 &  $-$62.6     &  $-$3.10$\pm$0.10 & $-$6.19$\pm$0.15    & 117.6$\pm$1.9 &  535.6$\pm$17.3 & 2008 October 3 \\
3 & IRS 28         &  $-$53.7    &  10.28$\pm$0.07 & $-$5.43$\pm$0.39     &  84.6$\pm$3.0  &  488.6$\pm$13.9  & 2008 June 3 \\
  &                &  $-$54.1    &  10.53$\pm$0.12 & $-$5.42$\pm$0.21     &  129.7$\pm$1.4  &  458.8$\pm$17.3 & 2008 October 3  \\
4 & SiO-15         & $-$35.5      & $-$12.38$\pm$0.15 &  $-$10.40$\pm$0.40     & 49.9$\pm$1.0  & 121.8$\pm$13.9  & 2008 June 3   \\
  &                & $-$36.0      & $-$12.08$\pm$0.13  &  $-$11.20$\pm$0.10      & 19.0$\pm$0.3  & 46.1$\pm$10.0  & 2008 October 3   \\
5 & SiO-19$^{*}$    &  $-$29.4    &  16.09$\pm$0.11  & $-$21.41$\pm$0.42    & 37.2$\pm$2.0   & 81.7$\pm$11.3   & 2008 June 3  \\
6 &  IRS 10EE       &  $-$26.3     & 7.62$\pm$0.03  &  4.74$\pm$0.12   &  572.8$\pm$10.0  &  2358.0$\pm$16.0  & 2008 June 3  \\
  &                 &  $-$26.6     & 7.89$\pm$0.06  &  4.75$\pm$0.08   &  472.8$\pm$2.5  &  1982.2$\pm$20.0 & 2008 October 3  \\
7 &  SiO-20$^{*}$    & $-$20.3   & $-$13.94$\pm$0.03  & 20.91$\pm$0.57    & 92.4$\pm$1.0  &  225.2$\pm$13.9   & 2008 June 3  \\
8 &  IRS 14NE$^{*}$    & $-$11.7    & 0.86$\pm$0.11  & $-$8.01$\pm$0.51    &  51.0$\pm$1.0  & 160.8$\pm$13.9   & 2008 June 3   \\
9 &  IRS 15NE         & $-$13.0    &  0.99$\pm$0.15 &  11.16$\pm$0.40   &   46.2$\pm$1.0    &  269.7$\pm$16.0   & 2008 June 3  \\
  &                   & $-$11.5    &  1.32$\pm$0.53 &  11.91$\pm$0.18   &   80.6$\pm$1.6  &  298.5$\pm$17.0  & 2008 October 3  \\
10 & SiO-16          &  + 9.1    & $-$26.54$\pm$0.05 &  $-$33.72$\pm$0.27    &  109.3$\pm$3.0  & 2828.0$\pm$17.9 & 2008 June 3   \\
   &                 &  + 7.7    & $-$26.33$\pm$0.22  &  $-$33.88$\pm$0.20    &  54.6$\pm$1.1  & 393.9$\pm$20.0  & 2008 October 3  \\
11 & SiO-21$^{*}$    &  + 13.8     & 40.74$\pm$0.08  &  $-$21.88$\pm$0.40   & 49.5$\pm$1.3  & 137.0$\pm$17.3  & 2008 June 3 \\
12 & SiO-22$^{*}$    &   + 33.6    &  41.28$\pm$0.06 &  15.63$\pm$0.22   &  265.8$\pm$5.0  &  1096.0$\pm$16.0 & 2008 June 3  \\
   &                 &   + 33.4    &  41.12$\pm$0.26 &  15.80$\pm$0.23   &  109.0$\pm$1.2  &  337.7$\pm$20.0   & 2008 October 3 \\
13  &  SiO-6           &   + 52.9    &  35.13$\pm$0.07 &  31.68$\pm$0.34   & 111.8$\pm$6.0  &  2427.0$\pm$24.5  & 2008 June 3  \\
    &                  &   + 52.5    &  35.64$\pm$0.73 &  31.13$\pm$0.37   & 56.5$\pm$1.8  &  295.4$\pm$22.4   & 2008 October 3  \\
14  & SiO-17          &  + 54.1    & 8.00$\pm$0.05  & $-$27.23$\pm$0.21    & 595.7$\pm$2.0  &  1887.0$\pm$20.0 & 2008 June 3 \\
    &                 &  + 54.0    & 8.63$\pm$0.03   &$-$27.15$\pm$0.02     & 112.3$\pm$0.7  &  281.1$\pm$10.0  & 2008 October 3 \\
15  & SiO-11         &  + 71.5     & 1.66$\pm$0.05  & 40.70$\pm$0.15   & 501.3$\pm$7.0  &  1767.0$\pm$17.3   & 2008 June 3  \\
    &                &  + 71.1     & 1.93$\pm$0.12  & 40.66$\pm$0.07   & 240.1$\pm$1.6  &  996.8$\pm$17.3    & 2008 October 3  \\
16  & SiO-12        & + 80.0      & $-$19.16$\pm$0.05  & 43.46$\pm$0.27     &  43.3$\pm$1.4   &  116.5$\pm$8.0  & 2008 June 3   \\
    &               & + 80.0      & $-$19.05$\pm$0.17  & 43.0790$\pm$0.12      &  33.0$\pm$1.2   &  79.1$\pm$10.0   & 2008 October 3   \\
17 & SiO-23$^{*}$    &   + 80.0    &  23.97$\pm$0.37  &  20.29$\pm$0.19    &  26.6$\pm$1.4  &  65.8$\pm$10.0  & 2008 October 3  \\
18  & IRS 19NW      & + 85.9     & 14.37$\pm$0.21  &  $-$18.28$\pm$1.14  &  56.6$\pm$4.5  &  208.4$\pm$17.3 & 2008 June 3  \\
    &               & + 84.1     & 15.36$\pm$0.48  &  $-$17.23$\pm$0.21  &  61.6$\pm$0.5  &  278.1$\pm$17.3   & 2008 October 3 \\
 \enddata
  \tablenotetext{*}{:new detection}
 \end{deluxetable}

\begin{figure}[bp]
\begin{center}
\includegraphics[width=7.2in]{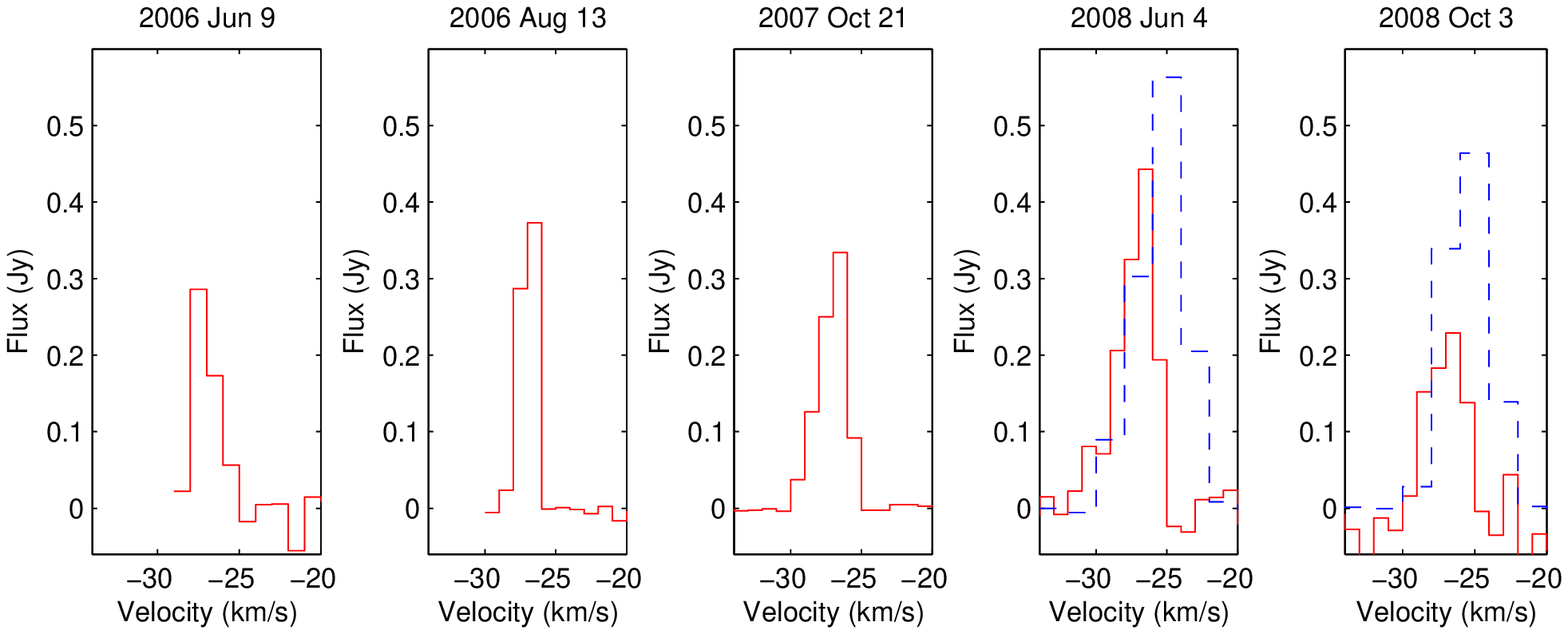}
\includegraphics[width=4.4in]{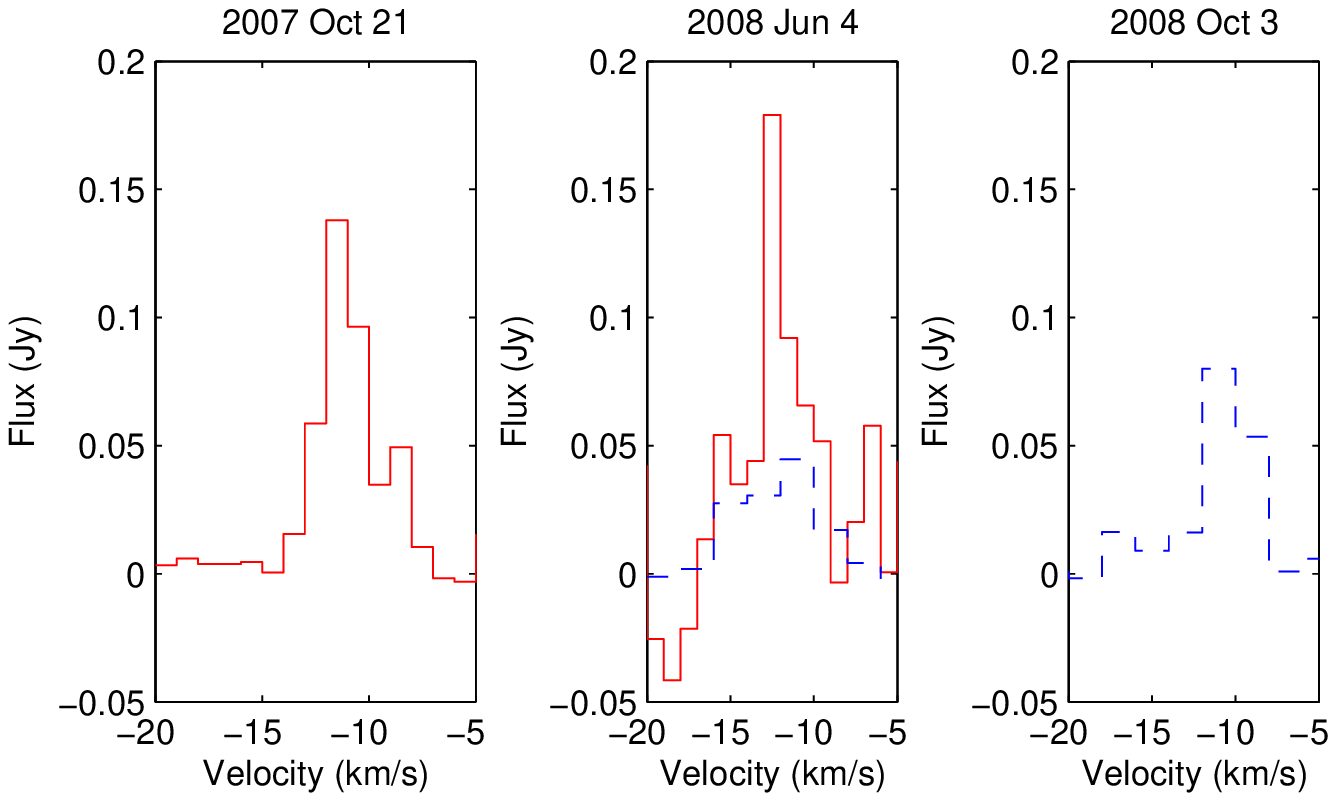}
\vspace*{-0.2 cm} \caption{SiO maser spectra of IRS 10EE ({\it upper
panel}) and IRS 15NE ({\it lower panel}) at 86 GHz (solid line) and
43 GHz (dashed line) from the ATCA observations. The velocity
resolutions is 1 km s$^{-1}$ at 86 GHz and 2 km s$^{-1}$ at 43 GHz.
The spectra were constructed at the pixel of peak brightness for SiO
masers. }
   \label{fig3}
\end{center}
\end{figure}

\begin{figure}[bp]
\begin{center}
\includegraphics[width=6.0in]{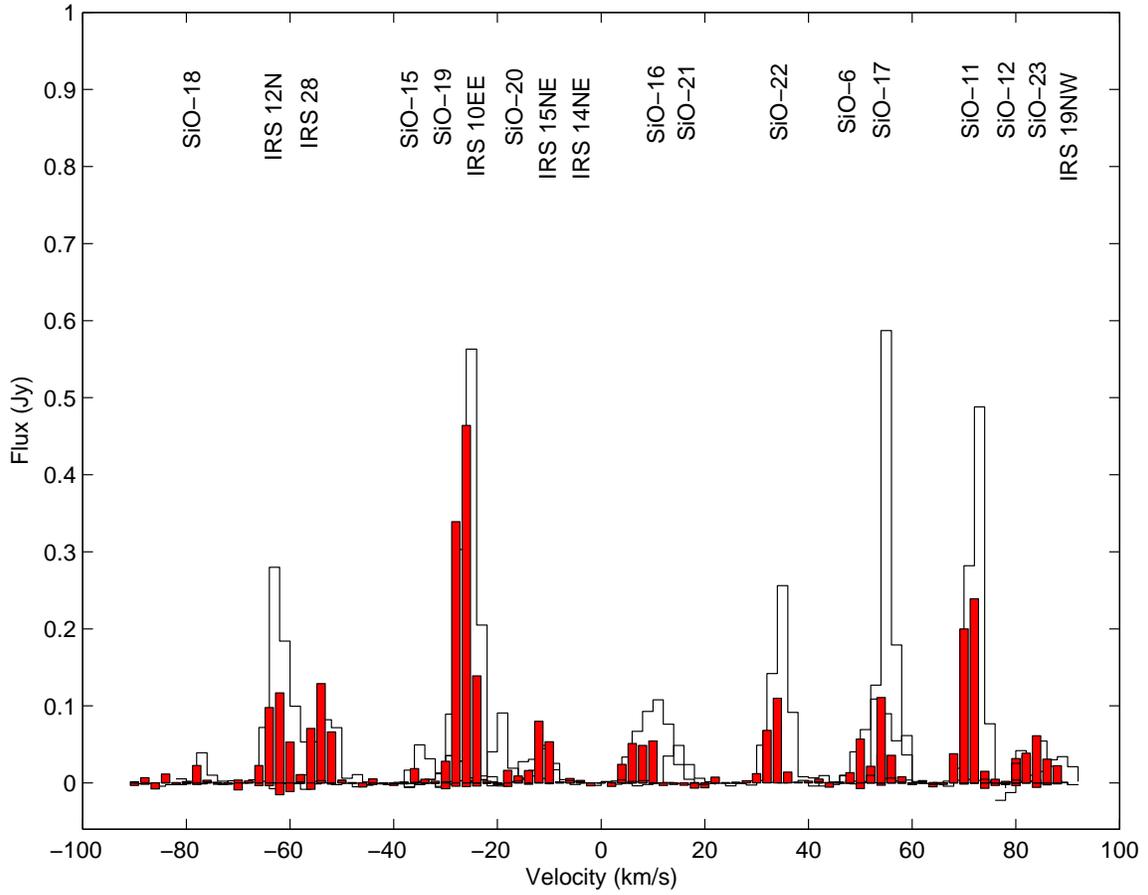}
\vspace*{-0.2 cm} \caption{Composite spectra of 43 GHz SiO masers
observed on 2008 June 3 (empty black bar) and October 3 (filled red
bar).}
\end{center}
\end{figure}

\begin{figure}[bp]
\begin{center}
\includegraphics[width=5.2in]{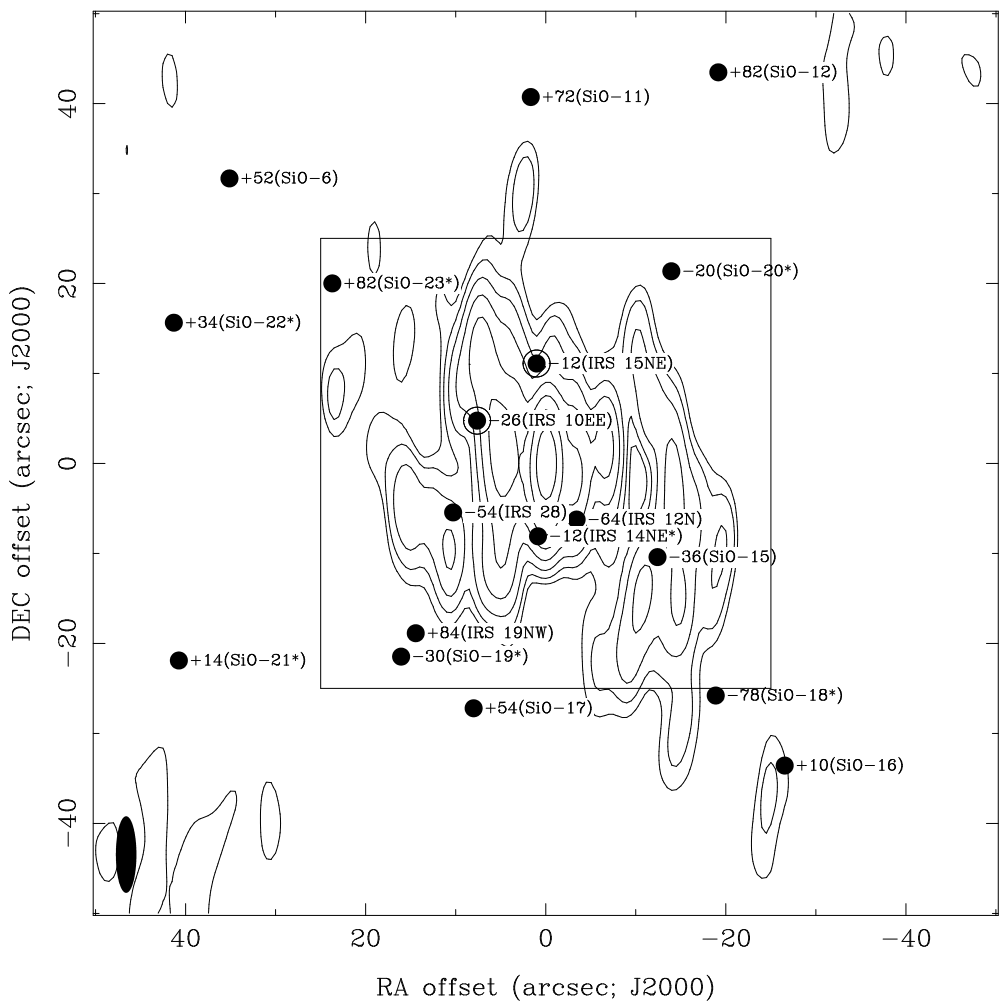}
\vspace*{-0.2 cm} \caption{The 7 mm ATCA continuum map of the GC
region observed on 2008 June 3 with a synthesized beam of
$8.^{\prime\prime}4~\times2.^{\prime\prime}3$ (shown at the bottom
left corner). The size of the region corresponds to an area of
projected size 4 pc $\times$ 4 pc centered on the position of Sgr
A*. Late-type giant and supergiant stars with eighteen 43 GHz SiO
maser (solid circle) emission and two 86 GHz SiO maser (open circle)
emission are marked. The seven new detections are marked with
*.
 }
   \label{fig2}
\end{center}
\end{figure}

\end{document}